\documentclass[preprint,12pt]{elsarticle}
\usepackage{amsfonts}
\usepackage{amsmath}
\usepackage{graphicx}
\usepackage{amssymb}

\biboptions{sort&compress}

\journal{Physica D}

\begin{document}
%\doublespacing

\newcommand{\dd}{d}
\newcommand{\pd}{\partial}
\newcommand{\myU}{\mathcal{U}}
\newcommand{\myr}{q}
\newcommand{\Urho}{U_{\rho}}
\newcommand{\myalpha}{\alpha_*}
\newcommand{\bd}[1]{\mathbf{#1}}
\newcommand{\Eq}[1]{Eq.\ (\ref{#1})}
\newcommand{\Eqn}[1]{Eq.\ (\ref{#1})}
\newcommand{\Eqns}[1]{Eqs.\ (\ref{#1})}
\newcommand{\myS}{Sec.\ }%{\S}
\newtheorem{theorem}{Theorem}

\begin{frontmatter}

%\preprint{Equilibrium and Stability}

\title{Continuum Modeling of the Equilibrium and Stability of Animal Flocks}% Force line breaks with \\

\author{Nicholas A. Mecholsky}
 \ead{nmech@umd.edu}
 \ead[url]{http://ape.umd.edu}
\author{Edward Ott}
\author{Thomas M. Antonsen, Jr.}
\author{Parvez Guzdar}
\address{
Institute for Research in Electronics \& Applied Physics,\\ University of Maryland, College Park, MD, 20742}
%University of Maryland, College Park, MD, 20742}
%\altaffiliation[Also at ]{Institute for Research in Electronics \& Applied Physics.}%Lines break automatically or can be forced with \\

%\date{8/25/10}
%\date{11/3/2010}
%\date{11/12/2010}
%\date{12/10/2010}%v_4
%\date{1/20/2011}%v_5
%\date{2/3/2011}%v_6
%\date{2/9/2011}%v_7
%\date{2/10/2011}%v_7a
%\date{3/9/2011}%v_8
%\date{3/17/2011}%v_9
%\date{3/18/2011}%v_10
%\date{3/21/2011}%v_11
%\date{3/23/2011}%v_12
%\date{5/23/2011}%v_2
%\date{8/12/2011}%v_3
%\date{8/15/2011}%v_4
%\date{8/16/2011}%v_5
\date{8/16/2011}%v_6

\begin{abstract}
Groups of animals often tend to arrange themselves in flocks that have characteristic spatial attributes and temporal dynamics.  Using a dynamic continuum model for a flock of individuals, we find equilibria of finite spatial extent where the density goes continuously to zero at a well-defined flock edge, and we discuss conditions on the model that allow for such solutions.  We also demonstrate conditions under which, as the flock size increases, the interior density in our equilibria tends to an approximately uniform value.  Motivated by observations of starling flocks that are relatively thin in a direction transverse to the direction of flight, we investigate the stability of infinite, planar-sheet flock equilibria.  We find that long-wavelength perturbations along the sheet are unstable for the class of models that we investigate.  This has the conjectured consequence that sheet-like flocks of arbitrarily large transverse extent relative to their thickness do not occur.  However, we also show that our model admits approximately sheet-like, `pancake-shaped', three-dimensional ellipsoidal equilibria with definite aspect ratios (transverse length-scale to flock thickness) determined by anisotropic perceptual/response characteristics of the flocking individuals, and we argue that these pancake-like equilibria are stable to the previously mentioned sheet instability.

%Flocks of birds and other aerial animals are observed in coordinated sheets of individuals that can move wildly in the air.  This paper investigates such sheet-like behavior.  Using a continuum model for a flock of individuals, we find equilibria where the density goes to zero at a finite extent for several different forms of the pressure.  We then investigate the stability of a sheet of individuals.  We find that motion of the flock along the direction of the sheet is always unstable for the pressures that we investigate.
\end{abstract}


\end{frontmatter}
%\maketitle

\section{Introduction}\label{sec:intro}
The formation and movement of groups of animals is a collective phenomenon emerging from behaviors of the individual group members.  One particularly striking example is the coordinated motion of large flocks of birds (such as starlings \cite{Ballerini2008201}).  The goal of this paper is to utilize a simple model to study equilibria and stability of flocks.  In particular, we note the following quote from Ref.\ \cite{Ballerini2008201} which reports an empirical investigation of starling flocks:
\begin{quote}
``Perhaps the most interesting morphological result is that flocks seemed to have a characteristic shape, being thin in the direction of gravity and more extended perpendicular to it. The proportions of the flock were well defined, with only weakly fluctuating aspect ratios, despite showing a wide range of sizes.  Our ability to conclude this stems entirely from the fact that we were able to analyze several groups with very different sizes (dimensions and number of birds).  Nonspherical shapes have also been observed in fish schools, the average proportions $I_1$:$I_2$:$I_3$ ranging from 1.0:1.7:2.1 in pilchards, \ldots 
% \emph{Harengula humeralis} (Cullen et al. 1965)
to 1.0:3.0:6.0 in saithes, \ldots
% \emph{Pollachius virens}, 
and 1:3:4 in herrings, \ldots.
%\emph{Clupea harengus} (Partridge et al. 1983).  
These values are comparable to those we found for starlings: 1.0:2.8:5.6.
\end{quote}
Analysis, to be presented here, shows that our simple model reproduce these features.  In particular, a flock tends to have the same aspect ratios independent of flock size.

There are two main modeling paradigms that have been used to investigate collective behavior of flocks.  In one, the position of each flock member and its interactions with other flock members and the environment are evolved \cite{Okubo86, Reynolds87, Vicsek95, Couzin02, Gregoire04, Shimoyama96, Flierl99, MikhailovZannette}, while the other is based on a continuum model evolving the density of individuals \cite{Topaz04, Levine00, TonerTu98PRE, Dorsogna06, Flierl99, Kulinski05, Mogilner99, Bertin06, Leverentz2009}.  In addition, within the class of continuum models, two subclasses may be distinguished which we call kinematic and dynamic (after Ref.\ \cite{Topaz04}).  In kinematic models the macroscopic flock velocity $\bd{v}(\bd{x},t)$ is instantaneously determined by the flock density, while dynamic models employ a velocity evolution equation specifying $\pd \bd{v}/\pd t$.  In what follows, we will use a dynamic continuum model.  A main motivation for our use of a continuum description is that continuum models are better suited to analyze stability and wave propagation.  For a review of this and other collective behavior phenomena, see Refs.\ \cite{ParrishandHammer, MikhailovandCalenbuhr, BonabeauDorigoTheraulaz, Camazine,HelbingReview01, TonerReview05}. 

This paper is organized as follows.  In \myS \ref{sec:modelFormulation} we present our continuum flock model.  In \myS \ref{sec:equilibria} we apply our continuum model to obtain planar, one-dimensional equilibrium solutions, as well as ellipsoidal three-dimensional equilibrium solutions.  We show that certain conditions on the model equations lead to finite flocks (i.e., zero density of flock members outside a finite spatial region) and others do not.  Similarly, we find that, if the model is further constrained, the flock's interior density does not blow up with increasing flock size (an unrealistic situation), but rather tends towards a well-defined, approximately uniform interior density.  In addition, we discuss how our results can be simply adapted to incorporate flock member interactions through the `topological distance' as advocated in Ref.\ \cite{Ballerini2008PNAS}\footnote{By topological distance, Ref.\ \cite{Ballerini2008PNAS} means that a flock member interacts with a fixed number of nearest flock members, independent of their geometrical distance.}.  In \myS \ref{sec:instability} we investigate an instability of a one-dimensional (`sheet') equilibrium to long-wavelength perturbations along the sheet.  Conclusions and discussion are given in \myS \ref{conclusions}.

\section{Model Formulation}\label{sec:modelFormulation}
\subsection{Continuum Model}
The equations for our model are similar to those of Ref.\ \cite{Mecholsky2010_OA},
\begin{align}\label{eqn:main1}
\frac{\partial \rho}{\partial t} + \nabla \cdot (\rho \bd{v}) &=0, \\\label{eqn:main2}
\frac{\pd \bd{v}}{\pd t} + \bd{v} \cdot \nabla \bd{v} &= -\frac{1}{\rho} \nabla P(\rho) - \nabla U + \frac{1}{\tau} \left(1 - \frac{v^2}{v_0^2} \right) \bd{v}.
\end{align}
In \Eq{eqn:main1}, $\rho(\bd{x},t)$ denotes the density of flock members at spatial point $\bd{x} = (x,y,z)$.  Equation (\ref{eqn:main1}) represents the conservation of the number of flock members.  The density $\rho (\bd{x},t)$ may be thought of as the number of flock members in a small region of dimension $l$ centered at the point $\bd{x}$ divided by the volume of the region.  An inherent assumption is that the flock is large enough that regions of size $l$ can contain many flock members yet still satisfy the condition that $l$ is substantially smaller than the macroscopic characteristic spatial scales of the flock.  The quantity $\bd{v} (\bd{x} ,t)$ denotes the macroscopic flock velocity at point $\bd{x}$ given by the average over a region of size $l$ centered at $\bd{x}$ of the individual velocities of flock members.  The quantity $P(\rho)$ will be referred to as the `pressure' due to its analogy with the fluid dynamical pressure and the similarity of the form of \Eq{eqn:main2} with the momentum equation of fluid dynamics.  The term $P(\rho)$ may be heuristically thought of as arising from dispersion, relative to $\bd{v} (\bd{x}, t)$, of the velocities of individuals within the regions of size $l$, referred to above, coupled with short-range repulsive behavior of individuals, where this short-range behavior is motivated by collision avoidance\footnote{As in Ref.\ \cite{Mecholsky2010_OA} an additional term, representing the tendency of flocking individuals to align their directions of motion, $\bd{v}/|\bd{v}|$, with those of their neighbors, can be included on the right-hand size of \Eq{eqn:main2}.  However, this term turns out to be identically zero for the equilibria  we investigate, and, for simplicity, we have thus omitted it.}.  In what follows, we shall investigate the consequences of assuming different forms for the function, $P(\rho)$.  The quantity $U(\bd{x},t)$ is a potential modeling the longer-range attractive behavior of individuals necessary to form a flock.  The last term in \Eq{eqn:main2} represents the tendency of flock members to have a preferred speed $v_0$ with respect to the medium through which the flock moves (e.g., air and water in the cases of flying birds and swimming fish), and $\tau$ is the rate of relaxation of $v (\bd{x}, t) = |\bd{v} (\bd{x},t)|$ to this preferred speed (see, for instance, Refs.\ \cite{Erdmann03, Erdmann02, Chuang07, Dorsogna06}).  Here we choose $U$ to be
\begin{equation}\label{eqn:Udef}
U = \int u(\bd{x}-\bd{x}') \rho(\bd{x}') \, \dd^{D} x', \\
%\bd{W} &= \int \rho(\bd{x}') w(\bd{x}-\bd{x}') (\bd{v}(\bd{x}')-\bd{v}(\bd{x})) \, \dd^{D} x',
\end{equation}
where $\dd^{D} x'$ represents a $D$-dimensional differential volume element ($D=1$ or 3 in this paper).  For convenience, the function $u(\bd{x})$ will be taken to have the particular form given by the solution of the equation,
\begin{equation}\label{eqn:littleu}
\nabla^2_{\rho} u(\bd{x}) - \kappa^2 u(\bd{x}) = u_0 \, \delta(\bd{x}),\\
%\nabla^2_{\nu} w(\bd{x}) - \kappa_{\nu}^2 w(\bd{x}) = w_0 \, \delta(\bd{x}),
\end{equation}
where,
\begin{align}\label{eqn:nabladef}
\nabla^2_{\rho} &= \nabla \cdot \mathbb{K} \cdot \nabla \\
\mathbb{K} &= \bd{s}_1 \bd{s}_1  \frac{\kappa^{2}}{\kappa^{2}_{1}}+\bd{s}_2 \bd{s}_2  \frac{\kappa^{2}}{\kappa^{2}_{2}}+\bd{s}_3 \bd{s}_3  \frac{\kappa^{2}}{\kappa^{2}_{3}},
\end{align}
and $\kappa$, $\kappa_{1, 2, 3}$, and $u_0$ are constants, and $\{ \bd{s}_1, \bd{s}_2, \bd{s}_3\}$ are a mutually orthogonal set of unit vectors.  Note that $\nabla^2_{\rho}$ is an anisotropic form of the Laplacian operator; $\nabla^2_{\rho} = \nabla^2$ for $\kappa^2 = \kappa^2_{1} = \kappa^2_{2} = \kappa^2_{3}$.  This form for $\nabla^2_{\rho}$ is introduced to allow for the natural anisotropy of behavior and visual perception of flocking individuals.  For example, in flying birds, $\bd{s}_1 (\bd{x}, t) = \bd{v}/v$, $\bd{s}_2$ is the average direction of orientation from wingtip to wingtip within regions of size $l$, and $\bd{s}_3 (\bd{x},t)$ is the average orientation of the direction perpendicular to the plane formed by the head and outstretched wings (i.e., $\bd{s}_3$ is typically in the direction of gravity).  For example, the solution of \Eq{eqn:littleu} for $u(\bd{x})$ in three dimensions ($D = 3$) is
\begin{equation}\label{eqn:littleuspecific}
u(\bd{x}) = \left[ u_0 / (4 \pi r) \right] \exp (- \kappa r), \quad r = \left\{ \bd{x} \cdot \mathbb{K}^{-1} \cdot \bd{x} \right\}^{1/2},
\end{equation}
where we have rejected the solution to \Eq{eqn:littleu} that blows up at $r \rightarrow \infty$.  Here, $\kappa^{-1}$ represents a chosen reference length scale for interactions between flock members.

We note that, while we have accounted for anisotropy in flock member behavior and perception, we have not attempted to account for forward/backward asymmetry.  In particular, one expects an individual to be more sensitive to conditions in front of it than to conditions at an equal distance behind it.  This type of forward/backward asymmetry is often accounted for in models that are based on equations of motion for each of many simulated individual flock members (for instance see Refs.\ \cite{Couzin02, Newman08}), but, so far, has not been incorporated in a continuum model.

\subsection{Other Continuum Models}\label{sec:continuummodels}
Continuum models have been used for many years to model collective behavior of animal groups \cite{TonerTu95PRL, TonerTu98PRE, Kulinski05, Chuang07, Mogilner99, Chen06, Czirok00, Erdmann02, Erdmann05, Leverentz2009, Topaz04, Topaz06, Mecholsky2010_OA, Flierl99, Levine00, Ramaswamy02}.  Most such models use continuity of agents to constrain the motion of individuals.  In addition to this, some researchers \cite{Topaz06, Leverentz2009, Topaz04, Mogilner99} set the velocity to be a particular function of the density and velocity and their spatial derivatives.  These models, which could be called kinematic (after Ref.\ \cite{Topaz04}), are to be distinguished from dynamic models \cite{TonerTu95PRL, TonerTu98PRE, Chuang07, Chen06, Mecholsky2010_OA, Flierl99, Levine00, Ramaswamy02} such as the model just presented.  In dynamic models the macroscopic flock velocity evolves in time through an equation for $\partial \bd{v}/\partial t$ (as in our momentum-like equation given in \Eq{eqn:main2}). 

Similar to discrete models, continuum models employ a subset of terms that represent the attraction, repulsion, orientation, self-propulsion, and noise.  Only a few continuum models have modeled the anisotropic behavior of sensing and response, most notably Refs.\ \cite{TonerReview05, TonerTu98PRE}.

Repulsion of nearby individuals is sometimes considered a nonlocal term \cite{Levine00, Chuang07}, analogous to the attractive potential $U$ given in \Eq{eqn:Udef}, but with the opposite sign and a shorter range.  Since the biologically relevant limit is for repulsion to have a shorter length-scale, we have directly incorporated repulsion through the local gradient of a `pressure' in \Eq{eqn:main2}; this is similar to Ref.\ \cite{TonerReview05}.  Attraction is longer range and is included in our model in a nonlocal manner (\Eqns{eqn:Udef} and (\ref{eqn:littleu}) for $U$ and $u$), and nonlocal modeling of attraction is also done in most other continuum and discrete flocking models.  Some authors have made fully local models (like Refs.\ \cite{Czirok00, TonerReview05, Chen06, Bertin06}), but these models cannot accurately simulate long-range attractive behavior.  Some researchers \cite{Mecholsky2010_OA, Leverentz2009, Chuang07, Topaz06, Topaz04, Levine00} use a nonlocal convolution of an exponential kernel and the density, similar to our model.  Others use attractive terms with kernels including piecewise continuous and power-law dependencies (see, for instance, Refs.\ \cite{Leverentz2009, Topaz06, Mogilner03}).  To what extent the form of these attraction terms affect the model and conclusions is not clear.

\section{Equilibria}\label{sec:equilibria}
We obtain solutions to our model equations that represent a uniformly translating flock.  To this end we set
\begin{equation}
\bd{v} = v_0 \bd{s},
\end{equation}
where $v_0$ and the unit vector $\bd{s}$ are constant everywhere within the flock.  After translating to a frame moving with the constant speed $\bd{v}$, \Eqns{eqn:main2}, (\ref{eqn:Udef}), and (\ref{eqn:littleu}) yield the equilibrium equations,
\begin{align}\label{eqn:forcebalance}
-\frac{1}{\rho} \nabla P(\rho) - \nabla U &= 0,\\ \label{eqn:bigU}
\nabla^2_{\rho} U - \kappa^2 U &= u_0 \rho,
\end{align}
where \Eq{eqn:bigU} results from \Eq{eqn:Udef} by replacing $\bd{x}$ by $\bd{x} - \bd{x}'$, multiplying by $\rho(\bd{x}')$ and then integrating the result over $\bd{x}'$.  Note that \Eq{eqn:forcebalance} applies only inside the flock ($\rho(\bd{x}) > 0$), since \Eq{eqn:main2} is irrelevant for $\rho = 0$ (outside a flock).  Equation (\ref{eqn:forcebalance}) can be rewritten
\begin{equation}
\nabla \left[ f(\rho) + U \right] = 0,
\end{equation}
where
\begin{equation}\label{eqn:fdef}
f(\rho) = \int_{0}^{\rho} \frac{1}{\rho'} \, \dd P(\rho').
\end{equation}
Thus within the flock
\begin{equation}\label{eqn:UandUb}
U(\bd{x}) + f(\rho) = U_{\textrm{B}},
\end{equation}
where $U_{\textrm{B}}$ is a constant.  Since we seek equilibria with $\rho = 0$ on the flock boundary, we impose $f=0$ and $U = U_{\textrm{B}}$ on the boundary.  In principle we can solve \Eq{eqn:UandUb} for $\rho$, yielding a result of the form
\begin{equation}\label{eqn:generalrho}
\rho = F(U_{\textrm{B}}-U),
\end{equation}
where
\begin{equation}\label{eqn:generalrho2}
f[F(x)] = x.
\end{equation}
Combining \Eqns{eqn:generalrho} and (\ref{eqn:bigU}) yields a nonlinear modified Helmholtz equation for $U$,
\begin{equation}\label{eqn:FULLU}
\nabla^2_{\rho} U - \kappa^2 U = u_0 \left\{ \begin{array}{cl}
F(U_{\textrm{B}}-U) & \textrm{inside the flock,} \\
0 & \textrm{outside the flock.} \end{array} \right.
\end{equation}
Integration of \Eq{eqn:FULLU} across the flock boundary yields the condition that $U$ and $\nabla U$ are continuous across the flock boundary.  Outside the flock ($\rho = 0$) the magnitude of the relevant solution of \Eq{eqn:FULLU} decays monotonically to zero at infinity in accordance with \Eqns{eqn:littleuspecific} and (\ref{eqn:Udef}).  Also, the potential must be confining within the flock to balance the repulsive action of the pressure.  These facts imply that $U$ is negative everywhere. 

We will consider a planar (`sheet') and three-dimensional ellipsoidal (`pancake') equilibria.  In the case of `sheet' equilibria, we assume a one-dimensional density profile, and thus we refer to this as a planar one-dimensional solution.  In this case $U$ depends only on $x$ and \Eq{eqn:FULLU} reduces to
\begin{equation}\label{eqn:1DU}
\xi^2 \frac{\dd^2 U}{\dd x^2} - \kappa^2 U = u_0 \left\{ \begin{array}{cl}
F(U_{\textrm{B}} - U) & \textrm{inside the flock,} \\
0 & \textrm{outside the flock,} \end{array} \right.
\end{equation}
where $\xi^2 = \hat{\bd{x}} \cdot \mathbb{K} \cdot \hat{\bd{x}} = \sum_{n=1}^{3} (\bd{s}_n \cdot \hat{\bd{x}})^2 (\kappa/\kappa_{n})^2$.  In the case of three-dimensional ellipsoidal equilibria, without loss of generality, we can take $\bd{s}_1 = \hat{\bd{x}}$, $\bd{s}_2 = \hat{\bd{y}}$, and $\bd{s}_3 = \hat{\bd{z}}$.  We further assume that $U$ is a function of the single variable
\begin{equation}\label{eqn:rdef}
r = \left\{ \bd{x} \cdot \mathbb{K}^{-1} \cdot \bd{x} \right\}^{1/2}=\left\{ \left( \kappa_{1} x \right)^2 + \left( \kappa_{2} y \right)^2 + \left( \kappa_{3} z \right)^2 \right\}^{1/2}/\kappa.
\end{equation}
In this case \Eq{eqn:FULLU} reduces to 
\begin{equation}\label{eqn:3dU}
\frac{1}{r^2} \frac{\dd}{\dd r} r^2 \frac{\dd U (r)}{\dd r} - \kappa^2 U(r) = u_0 \left\{ \begin{array}{cl}
F(U_{\textrm{B}}-U) & \textrm{inside the flock,} \\
0 & \textrm{outside the flock.} \end{array} \right.
\end{equation}

\subsection{Finite Flock Condition}\label{sec:flockcond}
We now consider the effects of different choices for the `equation of state', $P = P (\rho)$.  A main conclusion will be that if, as $\rho \rightarrow 0$, the pressure $P(\rho)$ behaves like
\begin{equation}\label{eqn:P}
P(\rho) \cong c \rho^{\gamma},
\end{equation}
then finite equilibria are possible only if 
\begin{equation}\label{eqn:gammacond}
\gamma > 1.
\end{equation}
Here by `finite equilibria' we mean equilibria in which $\rho > 0$ is continuous in a region $\mathcal{R}$, going to $\rho = 0$ on the region boundary $\partial \mathcal{R}$, and $\rho = 0$ outside $\mathcal{R}$.  Equation (\ref{eqn:gammacond}) follows as a consequence of the fact that \Eq{eqn:fdef} implies $f(\rho) \sim \rho^{\gamma - 1}$ near $\rho = 0$, and thus $f \rightarrow 0$ as $\rho \rightarrow 0^{+}$ only if \Eq{eqn:gammacond} holds.

\subsection{Behavior at the Flock Edge}\label{sec:flockedge}
\begin{figure}[!hb]
	\begin{center}
  	% Requires \usepackage{graphicx}
 	\includegraphics[width=14cm]{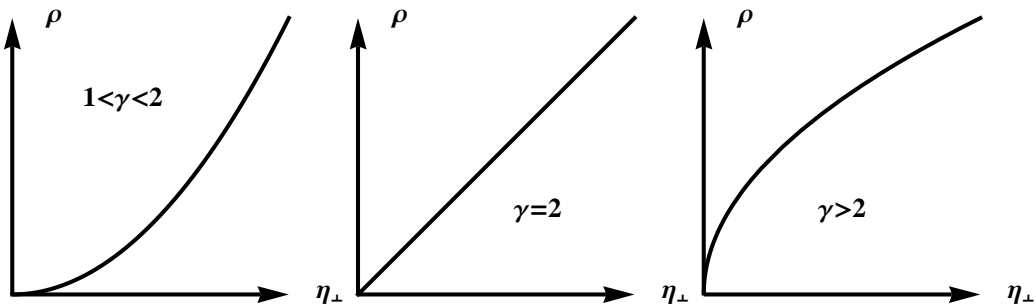}\\
      \caption{\label{fig:1}Schematic of the density dependence near the edge of the flock for different $\gamma$.}
	\end{center}
\end{figure}

Let $\eta_{\perp}$ denote a coordinate measuring the perpendicular distance to the flock edge, where $\eta_{\perp} > 0$ corresponds to points inside the flock.  Expanding \Eq{eqn:forcebalance} about a point $p$ on the flock boundary and assuming \Eq{eqn:P} for small $\rho$, \Eq{eqn:forcebalance} yields
\begin{equation}
\frac{1}{\rho} \frac{\dd \rho^{\gamma}}{\dd \eta_{\perp}} \cong J, \quad J = -\frac{1}{c} \frac{\dd U}{\dd \eta_{\perp}} \bigg|_{p},
\end{equation}
and the dependence of $\rho$ near the edge ($\eta_{\perp}$ small) is
\begin{equation}\label{eqn:edgerho}
\rho \sim \eta_{\perp}^{\frac{1}{\gamma - 1}},
\end{equation}
as illustrated schematically in Fig.\ \ref{fig:1}.

\subsection{The case, $P(\rho) = c \rho^2$}\label{sec:quadraticpressure}
\begin{figure}[!hb]
	\begin{center}
  	% Requires \usepackage{graphicx}
 	\includegraphics[width=9cm]{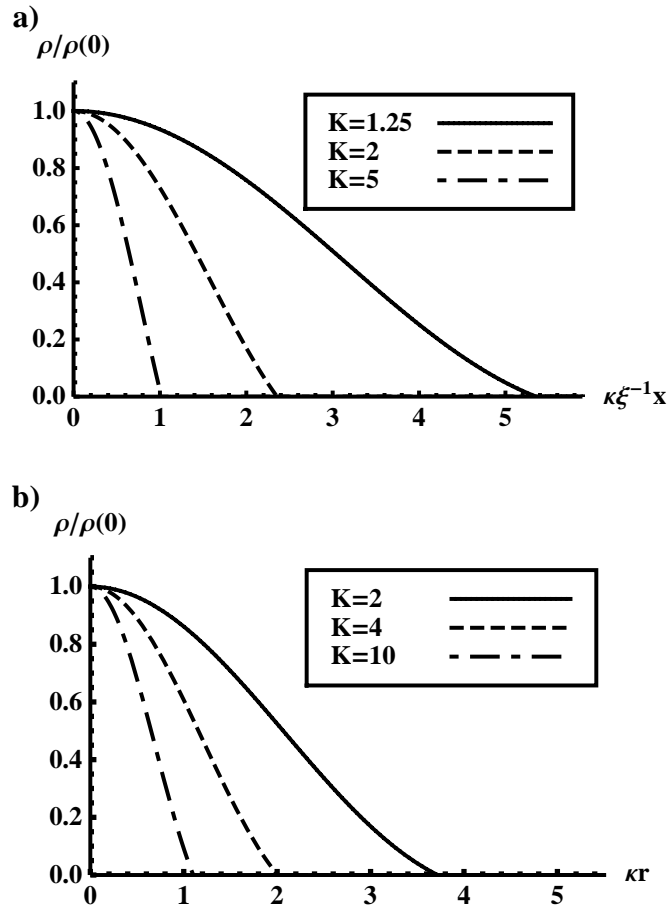}\\
      \caption{\label{fig:2}Density profiles for planar (a) and ellipsoidal (b) flock equilibria with $P=c \rho^2$ and $K = u_0/2 c \kappa^2$.}
	\end{center}
\end{figure}

In this case, \Eq{eqn:fdef} gives $f(\rho) = 2 c \rho$, which, when combined with \Eqns{eqn:UandUb} and (\ref{eqn:generalrho}), yields
\begin{equation}
F(U_{\textrm{B}} - U) = \left[ U_{\textrm{B}} - U(\bd{x})\right] / 2 c.
\end{equation}
Thus, for this special pressure dependence, \Eq{eqn:FULLU} is linear in $U(\bd{x})$, and this facilitates an analytical solution of the problem.  In the one-dimensional planar case, the solution of \Eq{eqn:1DU} within the flock ($|x| < x_{\textrm{B}}$) is
\begin{equation}\label{eqn:den}
\frac{\rho(x)}{\rho(0)} = \frac{\cos (\sqrt{K-1} \, \kappa \xi^{-1} x) - \cos (\sqrt{K-1}  \, \kappa \xi^{-1} x_{\textrm{B}})}{1 - \cos (\sqrt{K-1}  \, \kappa \xi^{-1} x_{\textrm{B}})},
\end{equation}
with $K = u_0 / (2 c \kappa^2)$.  Outside the flock ($|x| > x_{\textrm{B}}$ and $\rho = 0$), the solution for \Eq{eqn:1DU} is a negative exponential, $U(x) = U_{\textrm{B}} \exp ( - \kappa \xi^{-1} |x - x_{\textrm{B}}|)$.  Use of the conditions at the boundary (that $U$ is continuous and has a continuous derivative) determines the unknown $x_{\textrm{B}}$ in terms of the physical parameters $u_0$ and $c$,
\begin{equation}
\sin (\sqrt{K-1}  \, \kappa \xi^{-1} x_{\textrm{B}}) = -\sqrt{K - 1} \, \, \cos (\sqrt{K-1}  \, \kappa \xi^{-1} x_{\textrm{B}}).
\end{equation}
Several solutions of this type are shown in Fig.\ \ref{fig:2}(a). 

In the three-dimensional ellipsoidal case, the solution to \Eq{eqn:3dU} is 
\begin{equation}
\frac{\rho(r)}{\rho(0)} = \left[ \frac{\kappa r_{\textrm{B}} \sin \left( \sqrt{K - 1} \, \kappa r \right)}{\kappa r \sin \left( \sqrt{K - 1} \, \kappa r_{\textrm{B}} \right)} - 1 \right] \Bigg/ \left[\frac{\sqrt{K - 1} \, \kappa r_{\textrm{B}}}{\sin \left( \sqrt{K - 1} \, \kappa r_{\textrm{B}} \right)} - 1 \right].
\end{equation}
Outside ($r > r_{\textrm{B}}$), the density is zero and the boundary conditions give the boundary position as 
\begin{equation}
\sin( \sqrt{K - 1} \, \kappa r_\textrm{B}) = - \sqrt{K - 1} \, \, \cos( \sqrt{K - 1} \, \kappa r_\textrm{B} ) \, \, \left( 1 - K - \frac{K}{\kappa r_\textrm{B}}\right)^{-1}.
\end{equation}
Solutions of this type are shown in Fig.\ \ref{fig:2}(b) for various values of $K$.

Note that, as a consequence of the linearity of \Eqns{eqn:1DU} and (\ref{eqn:3dU}), for $\gamma = 2$, the density profiles of $\rho(x)/\rho(0)$ and $\rho(r)/\rho(0)$ and the normalized boundary location, $\kappa \xi^{-1} x_{\textrm{B}}$ and $\kappa r_{\textrm{B}}$, are determined by the model parameter $K$, and are independent of $\rho(0)$ (e.g., this is not so for $P(\rho) = c \rho^{\gamma}$ with $\gamma \neq 2$).

\subsection{One-Dimensional Equilibria}\label{sec:1D}

For a general pressure, $P(\rho)$, the solution in one planar dimension can be formally obtained by quadrature.  Multiplying \Eqn{eqn:1DU} by $\dd U/\dd x$, \Eqn{eqn:1DU} can be written as an exact differential.  Integrating this once yields a separable, first-order differential equation which, using \Eq{eqn:generalrho}, can be integrated to give
\begin{equation}\label{eqn:fullsolution}
x_{\textrm{B}} - x = \xi \int_{U(x)}^{U_{\textrm{B}}} \frac{\dd U}{\sqrt{\kappa^{2} U^2 - 2 u_0 P \left[ F(U_{\textrm{B}}-U) \right] }},
\end{equation}
where 
\begin{equation}
x_{\textrm{B}} = \xi \int_{U_{\textrm{min}}}^{U_{\textrm{B}}} \frac{\dd U}{\sqrt{\kappa^{2} U^2 - 2 u_0 P \left[ F(U_{\textrm{B}} - U) \right] }},
\end{equation}
and $U_{\textrm{min}} = U(0)$ is the root of 
\begin{equation}\label{Umin}
\kappa^{2} U_{\textrm{min}}^2 - 2 u_0 P \left[ F(U_{\textrm{B}} - U_{\textrm{min}}) \right]  = 0,
\end{equation}
such that the integrand is positive between $U_{\textrm{min}}$ and $U_{\textrm{B}}$.  Here $U_{\textrm{B}}$ is the value of the potential at the boundary of the flock and may be considered to be determined by specifying the number $N$ of individuals per unit transverse area in the flock,
\begin{equation}
N = \int_{-\infty}^{+\infty} \rho(x) \, \dd x.
\end{equation}
Since the potential $U$ is attractive, $U$ and $U_{\textrm{B}}$ are negative, and $U_{\textrm{B}}$ is the maximum value of $U$.  

We now consider pressures of the form $P(\rho) = c \rho^{\gamma}$ with $\gamma \neq 2$.  For $\gamma \leq 1$, we can show that there are no solutions where $\rho (x) \ge 0$ and $\int_{-\infty}^{\infty} \rho(x) \dd x < \infty$, and we therefore only consider $\gamma > 1$.  Using \Eqns{eqn:fdef} and (\ref{eqn:UandUb}) we obtain
\begin{equation}
F(U_{\textrm{B}} - U) = \rho = \left[ \frac{\gamma - 1}{c \gamma} \left( U_{\textrm{B}} - U \right) \right]^{\frac{1}{\gamma - 1}}.
\end{equation}
\begin{figure}[!hb]
	\begin{center}
  	% Requires \usepackage{graphicx}
 	\includegraphics[width=8cm]{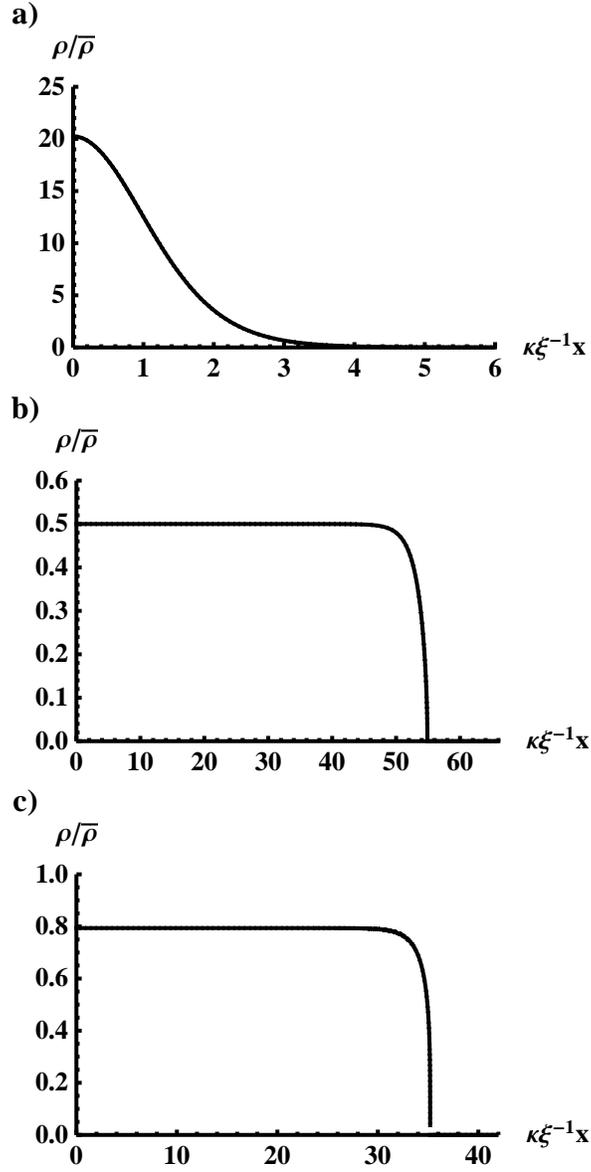}\\
      \caption{\label{fig:3} 1D Equilibrium density profiles for different pressures.  Density profile for a pressure of $P= c \rho^{\gamma}$ for a) $\gamma = 1.5$, b) $\gamma = 3.0$, and c) $\gamma = 5.0$.  Each profile has the same value for $\kappa N / (\bar{\rho} \xi) \approx 54$.}
	\end{center}
\end{figure}

Figure \ref{fig:3} shows density profiles for $\gamma = 1.5$, 3, and 5.0 for $P(\rho) = c \rho^{\gamma}$.  In Fig.\ \ref{fig:3}, 
\begin{equation}
\bar{\rho} = \left( \frac{u_0}{\kappa^2 c} \right)^{\frac{1}{\gamma - 2}}.
\end{equation}
These three plots all have the same value of $\kappa N / (\bar{\rho} \xi)$.  Note that the results for $\gamma = 3$ and $\gamma = 5$ are qualitatively similar and show that $\rho(x)$ rises rapidly as one moves into the flock from its boundary ($x = x_{\textrm{B}}$), becoming approximately constant in the interior of the flock.  In contrast, the result for $\gamma = 1.5$ is quite different, being peaked at $x= 0$, with a very much larger value of $\rho(0)$ and  a very much smaller value of $x_{\textrm{B}}$.

\begin{figure}[!hb]
	\begin{center}
  	% Requires \usepackage{graphicx}
 	\includegraphics[width=9cm]{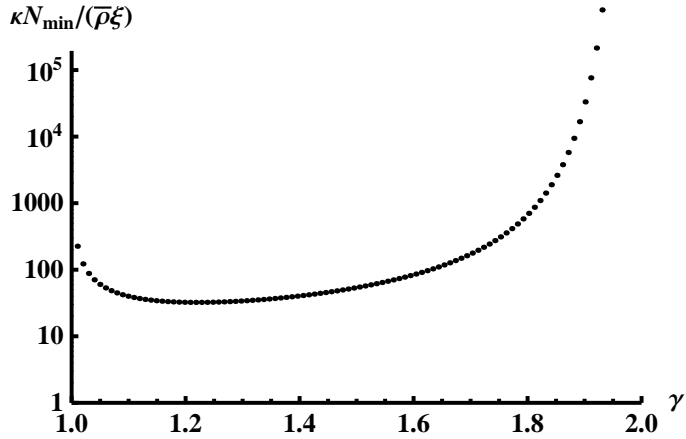}\\
      \caption{\label{fig:4} This figure shows the minimum number of individuals that will support a finite flock for the case of a pressure $P(\rho) = c \rho^{\gamma}$ for $1 < \gamma < 2$.}
	\end{center}
\end{figure}
When $1 < \gamma < 2$, there is a minimum number of individuals, parametrized by $\kappa_{\rho} N / (\bar{\rho} \xi)$, that will support a finite flock.  Fixing $U_{\textrm{B}}$ sets the total number of individuals in the flock.  A large $|U_{\textrm{B}}|$ corresponds to a large flock size.  The smallest $|U_{\textrm{B}}|$ can be is 0, which corresponds to a minimum flock size. This minimum value of $\kappa N / (\bar{\rho} \xi)$ is plotted as a function of $\gamma$ in Fig.\ \ref{fig:4}.  We see from the figure that the minimum flock number approaches infinity as we approach $\gamma = 2$ from the left.

\begin{figure}[!hb]
	\begin{center}
  	% Requires \usepackage{graphicx}
 	\includegraphics[width=9cm]{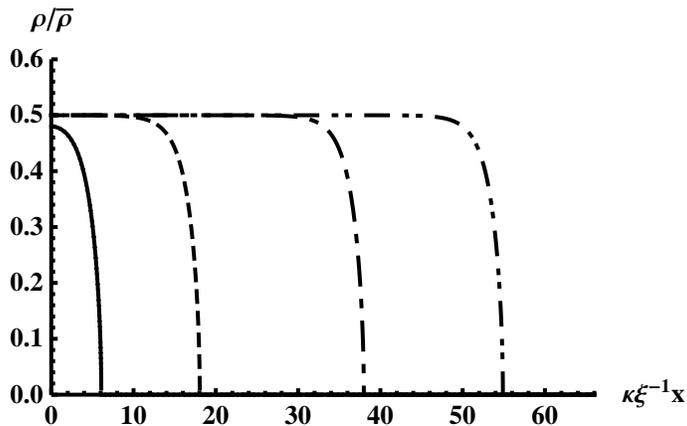}\\
      \caption{\label{fig:5} Density profiles for the pressure $P(\rho) = c \rho^{3}$ for various $\kappa N / (\bar{\rho} \xi)$.  The curves represent $\kappa N / (\bar{\rho} \xi)$ values of 4.8, 16.8, 36.7, and the last, 54, is the same curve as Fig.\ \ref{fig:3}(b).}
	\end{center}
\end{figure}
In Fig.\ \ref{fig:5}, we show density profiles for different values of $\kappa N / (\bar{\rho} \xi)$ in the case $\gamma = 3$.  We see that, for small $\kappa N / (\bar{\rho} \xi)$, the density profile is not constant near the center of the flock ($x=0$).  However, as the flock number increases, we approach the flat internal density and sharp cutoff depicted in Fig.\ \ref{fig:3}(b).

\begin{figure}[!hb]
	\begin{center}
  	% Requires \usepackage{graphicx}
 	\includegraphics[width=13.7058cm]{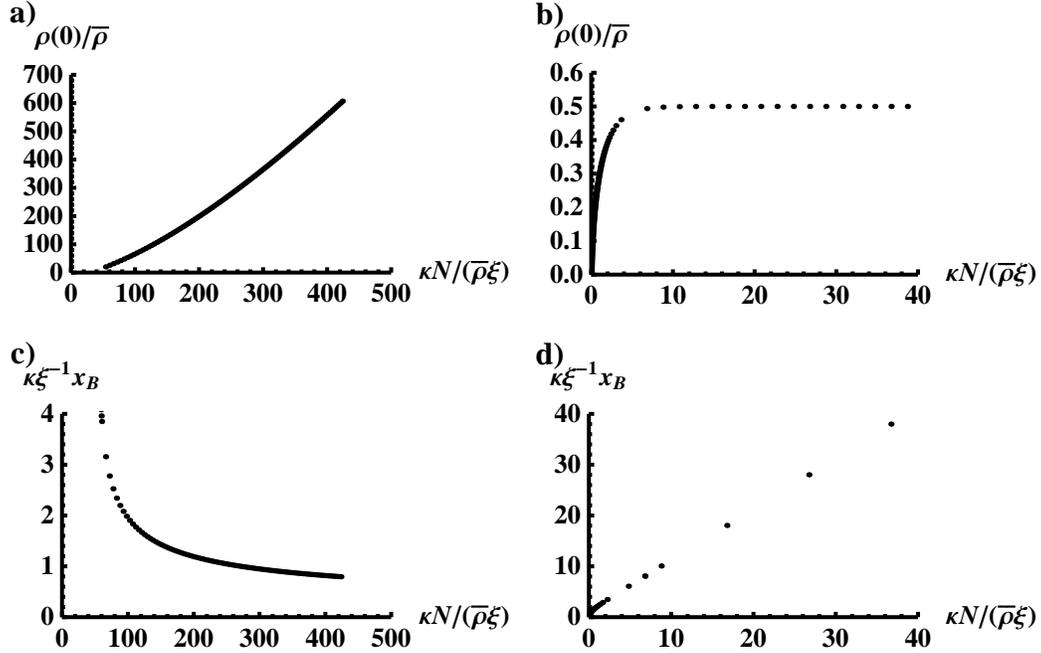}\\
      \caption{\label{fig:6} Maximum density, $\rho(0)/\bar{\rho}$, versus flock number, $\kappa N / (\bar{\rho} \xi)$, for a) $\gamma = 1.5$ and b) $\gamma = 3.0$.  Boundary position, $\kappa \xi^{-1} x_{\textrm{B}}$, versus flock number, $\kappa N / (\bar{\rho} \xi)$, for c) $\gamma = 1.5$ and d) $\gamma = 3.0$.}
	\end{center}
\end{figure}
\begin{figure}[!hb]
	\begin{center}
  	% Requires \usepackage{graphicx}
 	\includegraphics[width=9cm]{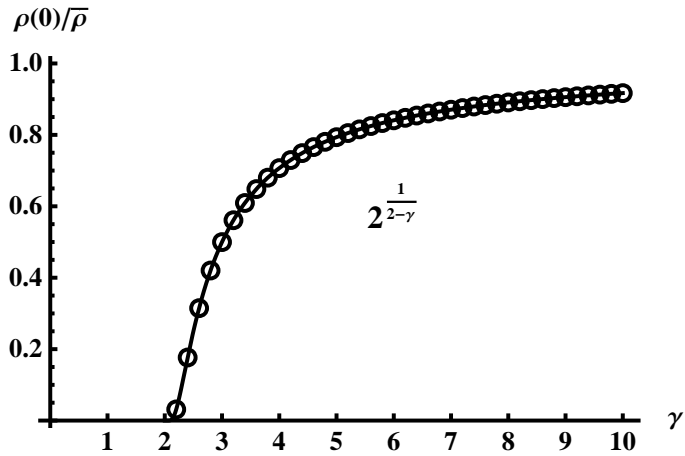}\\
      \caption{\label{fig:7} Plot of the density at $x=0$ versus $\gamma$ with a pressure of the form $P(\rho) = c \rho^{\gamma}$.  The black curve is the theoretical expression \Eq{eqn:centraldensitytheory}.  The open circles are the central densities of large flocks obtained by numerical integration of \Eq{eqn:fullsolution}.}
	\end{center}
\end{figure}

In Fig.\ \ref{fig:6} we plot the central flock density $\rho(0)/\bar{\rho}$ versus $\kappa N / (\bar{\rho} \xi)$ and the flock boundary $\kappa \xi^{-1} x_{\textrm{B}}$ versus $\kappa N / (\bar{\rho} \xi)$ for $\gamma = 1.5$ [Figs.\ \ref{fig:6}(a,c)] and $\gamma = 3.0$ [Figs.\ \ref{fig:6}(b,d)].  For $\gamma = 1.5$, we see in Fig.\ \ref{fig:6}(a) that the density at $x=0$, $\rho(0)$, increases faster than linear as $\kappa N / (\bar{\rho} \xi)$ increases.  In the case $\gamma = 2$ (described in \myS \ref{sec:quadraticpressure}), we found that $\rho(x)/\rho(0)$ and $x_{\textrm{B}}$ are independent of $\rho(0)$, implying that $\rho(0)$ is linearly proportional to $N$.  The faster-than-linear increase of $\rho(0)$ with $\kappa N / (\bar{\rho} \xi)$ is consistent with Fig.\ \ref{fig:6}(c) which shows that the position of the flock boundary, $x_{\textrm{B}}$, decreases with $\kappa N / (\bar{\rho} \xi)$.  For the case $\gamma = 3.0$, we see in Fig.\ \ref{fig:6}(b) that the density at the center of the flock saturates at a constant value as $\kappa N / (\bar{\rho} \xi)$ increases.  Consistent with this, Fig.\ \ref{fig:6}(d) shows the flock boundary position $x_{\textrm{B}}$ increasing proportionally to $\kappa N / (\bar{\rho} \xi)$ for large $\kappa N / (\bar{\rho} \xi)$.  Hence we conclude that for pressures of the form $P(\rho) = c \rho^{\gamma}$, with $\gamma = 3$, large flocks of different sizes will have the same flat interior density.  In fact we find this to be the case for any $\gamma > 2$.  Furthermore, for $\gamma > 2$ we can obtain an explicit expression for the central density, as follows.  From \Eq{eqn:1DU}, we see that, if the density is nearly constant in the interior of the flock, then $\dd^2 U/\dd x^2 \cong 0$ and
\begin{equation}\label{Rhomin}
\rho(0) \cong -\frac{\kappa^{2}}{u_0} U_{\textrm{min}}.
\end{equation}
From \Eq{Umin}
\begin{equation}
\kappa^{2} U_{\textrm{min}}^2 - 2 u_0 c \rho(0)^{\gamma}  = 0.
\end{equation}
Using \Eq{Rhomin} in \Eq{Umin} gives
\begin{equation}\label{eqn:centraldensitytheory}
\frac{\rho(0)}{\bar{\rho}} = 2^{-\frac{1}{\gamma - 2}}.
\end{equation}
Equation (\ref{eqn:centraldensitytheory}) is plotted as the solid curve in Fig.\ \ref{fig:7}; the open circles in Fig.\ \ref{fig:7} are the central density for a large flock evaluated by numerical integration of \Eq{eqn:fullsolution}.

As noted in the quote in \myS \ref{sec:intro}, it is common for animals of a given species (e.g., starlings) to form flocks with different numbers of flock members.  As the number of flock members increases the spatial extent of the flock also increases.  For a model with $P(\rho) = c \rho^{\gamma}$, our results above show that this type of observed behavior is only consistent with $\gamma > 2$.  Furthermore, it is expected that the interior density of larger and larger flocks eventually saturates, since individuals would be expected to become `uncomfortable' beyond a certain crowding density.  This is also consistent with $\gamma > 2$ since we obtain an interior density for large flocks, \Eq{eqn:centraldensitytheory}, that saturates with large $\kappa N / (\bar{\rho} \xi)$.

\subsection{Three-Dimensional Ellipsoidal Equilibria}\label{sec:3D}
Unlike our analysis of the planar one-dimensional case in \myS \ref{sec:1D}, the three-dimensional ellipsoidal problem, \Eq{eqn:3dU}, cannot be solved by the method of quadrature.  While one could solve \Eq{eqn:3dU} numerically, we feel that this is not necessary, as insight from the planar case can be used to give a good understanding of the ellipsoidal case.  In particular, we focus on the physically interesting case of $\gamma > 2$ and large flocks.  From our discussion in \myS \ref{sec:1D}, we expect that the interior density will again be flat, and, neglecting $r^{-2} \dd/\dd r (r^2 \dd U/\dd r)$ in the flat region, we again obtain \Eq{eqn:centraldensitytheory} for the interior density.  Furthermore, consider the thickness $\Delta r$ of the transition region, $r_{\textrm{B}} - \Delta r < r < r_{\textrm{B}}$, where $\rho(r)$ rises from $\rho(r_{\textrm{B}})=0$ (at the flock edge) to $\rho(r)$ near $\rho(0)$.  For large flocks, we expect that $\Delta r \ll r_{\textrm{B}}$.  Thus the effect of curvature of the flock boundary will have little effect on $\rho(r)$ within the transition region, and the spatial dependence of $\rho$ on $r$ from the boundary will be nearly the same as in the one-dimensional planar case.  Thus, for large flocks $\rho(r)$ will be the same as $\rho(x)$ given Figs.\ \ref{fig:3}(b and c) provided that we replace the horizontal axis variable $\kappa \xi^{-1} x$ in Figs.\ \ref{fig:3}(b and c) by $\kappa r$, and provided that the planar and ellipsoidal flock sizes are the same (in the sense that $x_{\textrm{B}} = r_{\textrm{B}}$).  Finally, replacing $r$ by \Eq{eqn:rdef} we obtain the pancake-shaped ellipsoidal flock density as a function of $(x, y, z)$.  Referring to the quote in \myS \ref{sec:intro}, the aspect ratios of the flock shape are $1$:$(\kappa_{2}/\kappa_{1})$:$(\kappa_{3}/\kappa_{1})$, and, in agreement with the observations, these aspect ratios for our model are independent of the flock size.

\subsection{`Topological Distance'}\label{sec:topology}
A recent paper \cite{Ballerini2008PNAS} suggests that the interactions between flocking starlings is through what they call the `topological distance', rather than the geometric distance.  Thus, for example, in the case of isotropic interactions, each bird may be thought of as interacting with a fixed number of other birds that are closest to it independent of the geometrical distance between them.  In contrast, our analysis above has assumed that interaction strength between individuals falls off with increasing geometrical distance.  In this subsection we show how our analysis above may be simply adapted to the case where flocking individuals interact through topological distance.  We accomplish this by replacing the operator $\nabla_{\rho}^2$, defined by \Eq{eqn:nabladef}, by the modified form
\begin{equation}
\hat{\nabla}_{\rho}^2 = \frac{\rho_{*}}{\rho} \nabla \cdot \left( \mathbb{K} \frac{\rho_{*}}{\rho} \right) \cdot \nabla,
\end{equation}
where $\rho$ is the flock density, and $\rho_{*}$ is a somewhat arbitrary reference density introduced to leave the units of $\hat{\nabla}_{\rho}^2$ as the inverse of length squared.  A convenient choice for $\rho_{*}$ might be the density in the center of the flock.  We now consider the effect of this modification in the case of a one-dimensional planar flock for which our formulation using geometrical distance yields \Eq{eqn:1DU}.  With our modification, \Eq{eqn:1DU} still applies, but with $x$ replaced by 
\begin{equation}
\tilde{x} (x) = \int_{0}^{x}  \frac{\rho (x ' )}{\rho_{*}}  \, \dd x',
\end{equation}
which is essentially the topological distance, as described in Ref.\ \cite{Ballerini2008PNAS}.  Considering a large flock, the central density $\rho(0)$ is thus still given by our previous analysis, \Eq{eqn:centraldensitytheory}, and the density for large flocks is again approximately constant in the flock interior.  To find the dependence of $\rho$ on distance from the flock edge we replace $\eta_{\perp}$ in \Eq{eqn:edgerho} by the corresponding topological quantity, $\tilde{\eta}_{\perp}$.  This yields
\begin{equation}\label{eqn:edgerho2}
\rho \sim \eta_{\perp}^{1/(\gamma - 2)},
\end{equation}
which is to be contrasted with the dependence $\rho \sim \eta_{\perp}^{1/(\gamma - 1)}$ in \Eq{eqn:edgerho}.  Note that in Sec.\ \ref{sec:1D} we have argued that $\gamma > 2$ for physical flocks with an assumed pressure of the form $P \sim \rho^{\gamma}$.  Thus, in this case, \Eq{eqn:edgerho2} continues to predict that $\rho$ goes continuously to zero at the flock edge.  Furthermore, by the discussion in Sec.\ \ref{sec:3D}, our consideration above for the one-dimensional planar case, directly translates to the case of an ellipsoidal flock.

\subsection{Other Continuous Equilibria}
In our approach we regard density profiles that go continuously to zero at the flock boundary and approach a constant uniform interior density at large flock size as `reasonable', and we have delineated conditions in our model class where this holds.  Furthermore, our work shows that our reasonable equilibria are robust in the sense that they exist in regions of parameter space (rather than just at a critical parameter value).

In contrast, Ref.\ \cite{Levine00%, Leverentz2009, Topaz04, Topaz06
} presents equilibria where the density profile goes discontinuously to zero at the flock edge, while another \cite{Flierl99} finds equilibria that give continuous density distributions that asymptote to a small constant density far from a localized flock.  References \cite{Leverentz2009, Topaz04, Topaz06} obtain profiles of the type that we regard as reasonable; they, however, do so using a kinematic model rather than a dynamic model as used here (see \myS \ref{sec:continuummodels} for our definition of dynamic vs. kinematic).  Our flock solutions, in which the density is strictly zero outside the flock, is similar to `compacton' solutions of certain nonlinear PDEs \cite{Rosenau}.

We note that we have restricted our investigation of flock equilibria to the case of translating flocks.  In particular, we have not attempted to address milling flocks in which individuals travel roughly along circular paths creating a vortex-like flock pattern.  Some papers investigating this type of flock are Refs.\ \cite{Levine00, Chen06, Newman08, Couzin02, Erdmann02, Chuang07, Ratushnaya06, Newman08, Topaz04}.

Finally, we note that field data on starling flocks \cite{Ballerini2008201} show that the density of birds increases as the border of the flock is approached.  This behavior is not present in our solutions.  Thus our model would need modification to describe this feature of starling flocks.  Nevertheless, we believe that our model is still useful in general, and this may be more strongly the case for animal groups of other types than starling flocks, e.g., fish schools.

\section{Long-Wavelength Instability of Planar Equilibria}\label{sec:instability}
\begin{figure}[!hb]
	\begin{center}
  	% Requires \usepackage{graphicx}
 	\includegraphics[width=13.7058cm]{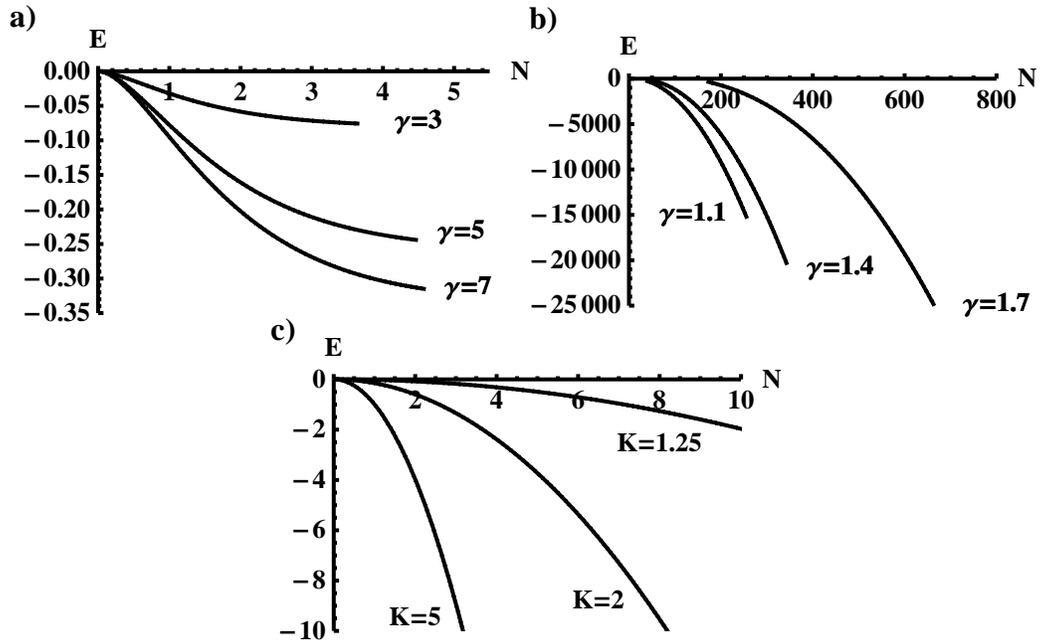}\\
      \caption{\label{fig:8} $E$ versus $N$ for various pressures. a) Numerical evaluation of $E$ versus $N$ for $P=\rho^{\gamma}$ for $\gamma > 2$.  b) Numerical evaluation of $E$ versus $N$ for $1 < \gamma < 2$.  c) Plot of the analytical result of $E$ versus $N$ for $P = \rho^2$.}
	\end{center}
\end{figure}

Here we analyze the stability of our planar one-dimensional equilibria obtained in \myS \ref{sec:equilibria} to long-wavelength perturbations.  For simplicity, we consider an equilibrium of individuals with constant flight velocity lying in the surface of the sheet, $\bd{v}_0 = v_0 \bd{z}_0$. We now introduce a perturbation depending on $x$ and $y$,
\begin{align}\label{eqn:ansatz1}
\rho(\bd{x}) &= \rho_0 (x) + \delta \rho(x) e^{s t + i k y}, \\ \label{eqn:ansatz2}
\bd{v}(\bd{x}) &= \bd{v}_0 + \delta \bd{v} (x)  e^{s t + i k y},
\end{align}
where the density profile $\rho_0 (x)$ is determined in \myS \ref{sec:1D}.  Using \Eqns{eqn:ansatz1} and (\ref{eqn:ansatz2}) in \Eqns{eqn:main1} and (\ref{eqn:main2}), we obtain
\begin{align}\label{eqn:dim_cont}
s \, \delta \rho &= - \nabla \cdot \left( \rho_0 \delta \bd{v} \right)\\ \label{eqn:dim_cont2}
s \, \delta \bd{v} &= -\frac{1}{\rho_0} \frac{\pd P}{\pd \rho} \nabla \left( \rho_0 + \delta \rho \right) - \nabla \left( U_0 + \delta U \right) - 2 \tau^{-1} \left( \bd{v}_0 \cdot \delta \bd{v} \right) \bd{v}_0.
\end{align}
For perturbations that are functions of $x$ and $y$, the perturbed velocity of the unstable mode will lie in the $x-y$ plane, and, since we take $\bd{v}_0 = v_0 \bd{z}_0$, we have that the last term on the right-hand side of \Eq{eqn:dim_cont2} is zero ($\bd{v}_0 \cdot \delta \bd{v} = 0$).  Integrating across the flock, \Eq{eqn:dim_cont} yields
\begin{equation}\label{eqn:perturbationeqn1}
s \int_{-x_{\textrm{B}}}^{x_{\textrm{B}}} \delta \rho \, \dd x = - i k \int_{-x_{\textrm{B}}}^{x_{\textrm{B}}} \rho_0 \delta v_y \, \dd x.
\end{equation}
The $y$ component of the velocity equation, \Eq{eqn:dim_cont2}, is
\begin{equation}\label{eqn:perturbationeqn2}
s \int_{-x_{\textrm{B}}}^{x_{\textrm{B}}} \rho_0 \delta v_y \, \dd x = - i k \int_{-x_{\textrm{B}}}^{x_{\textrm{B}}} \left[ \rho_0 \delta U + \frac{\pd P}{\pd \rho} \delta \rho \right] \, \dd x.
\end{equation}
Combining \Eqns{eqn:perturbationeqn1} and (\ref{eqn:perturbationeqn2}) we get
\begin{equation}\label{eqn:ssquared}
s^2 \int_{-x_{\textrm{B}}}^{x_{\textrm{B}}} \delta \rho \, \dd x = - k^2 \int_{-x_{\textrm{B}}}^{x_{\textrm{B}}} \left[ \rho_0 \delta U + \frac{\pd P}{\pd \rho} \delta \rho \right] \, \dd x.
\end{equation}
For long-wavelengths, $(k x_{\textrm{B}}) \ll 1$, an expansion in the small parameter $(k x_{\textrm{B}})$ shows that, to lowest order, at each point $y$ the equilibrium in $x$ (i.e., \Eq{eqn:1DU}) holds.  This is essentially equivalent to the statement that for $(k x_{\textrm{B}}) \ll 1$, derivatives in $x$ are much larger than derivatives in $y$ and $t$.  However, the integral,
\begin{equation}\label{eqn:Ndef}
N(y,t) = \int \rho \, \dd x,
\end{equation}
giving the flock number density per unit area transverse to $x$ can depend (slowly) on $y$ and $t$, and the quasi-static equilibrium depends on $y$ and $t$ only through $N$. Thus explicitly considering the equilibrium dependence on $N$ by writing $\rho_0$ and $U_0$ as $\rho_0 (x, N)$ and $U_0 (x, N)$, and setting $N = N_0 + \delta N$, we can express the perturbed quantities in \Eq{eqn:ssquared} as
\begin{align}\label{eqn:ansatzN1}
\delta \rho &= e^{s t - i k y} \frac{\pd \rho_0}{\pd N} \delta N,\\\label{eqn:ansatzN2}
\delta U &= e^{s t - i k y} \frac{\pd U_0}{\pd N} \delta N.
\end{align}
Using our definition of the cross-sectional number, \Eq{eqn:Ndef}, we obtain from \Eqns{eqn:ssquared}, (\ref{eqn:ansatzN1}), and (\ref{eqn:ansatzN2})
\begin{equation}\label{eqn:sequation}
s^2 \delta N = -k^2 \int_{-x_{\textrm{B}}}^{x_{\textrm{B}}} \left[ \rho_0 \frac{\pd U_0}{\pd N} + \frac{\pd P}{\pd \rho} \frac{\pd \rho_0 }{\pd N} \right] \delta N \, \dd x.
\end{equation}
We re-express the integral $\int \rho_0 (\partial U_0 / \partial N) \, \dd x$ as follows.  First we start with the dimensionless potential equation (see \Eq{eqn:1DU} with normalizations chosen to set $\xi^2 = \kappa^2 = u_0 = 1$),
\begin{equation}\label{eqn:dimPot}
\frac{\dd^2}{\dd x^2} U_0 - U_0 = \rho_0.
\end{equation}
We take a partial derivative of \Eq{eqn:dimPot} with respect to $N$, multiply by $U_0$, and integrate from $-x_{\textrm{B}}$ to $x_{\textrm{B}}$ (recall that the flock density goes to zero at the boundaries) giving
\begin{equation}\label{eqn:pot1}
\int_{-x_{\textrm{B}}}^{x_{\textrm{B}}} U_0 \left[ \frac{\dd^2 }{\dd x^2} - 1 \right] \frac{\pd U_0}{\pd N} \, \dd x = \int_{-x_{\textrm{B}}}^{x_{\textrm{B}}} U_0 \frac{\pd \rho_0}{\pd N} \, \dd x.
\end{equation}
We also multiply the original potential equation \Eqn{eqn:dimPot} by $\pd U_0 /\pd N$ and integrate from $-x_{\textrm{B}}$ to $x_{\textrm{B}}$, to obtain
\begin{equation}\label{eqn:pot2}
\int_{-x_{\textrm{B}}}^{x_{\textrm{B}}} \frac{\pd U_0}{\pd N} \left[ \frac{\dd^2 }{\dd x^2} - 1 \right] U_0 \, \dd x = \int_{-x_{\textrm{B}}}^{x_{\textrm{B}}} \frac{\pd U_0}{\pd N} \rho_0 \, \dd x.
\end{equation}
We then subtract \Eqn{eqn:pot1} from \Eqn{eqn:pot2}.  Using integration by parts twice cancels all terms on the left hand side.  There are no boundary terms in the integration by parts since $\rho_{0}( \pm x_{\textrm{B}}) = 0$.  We then have
\begin{equation}
\int U_0 \frac{\pd \rho_0}{\pd N} \, \dd x = \int \frac{\pd U_0}{\pd N} \rho_0 \, \dd x.
\end{equation}
Thus
\begin{equation}
\int_{-x_{\textrm{B}}}^{x_{\textrm{B}}} \rho_0 \frac{\pd U_0}{\pd N} \, \dd x = \frac{1}{2} \frac{\pd }{\pd N} \int_{-x_{\textrm{B}}}^{x_{\textrm{B}}} \rho_0 U_0 \, \dd x.
\end{equation}
Using this in \Eqn{eqn:sequation} gives the final result,
\begin{equation}\label{eqn:s}
s^2 \delta N = -k^2 \delta N \frac{\pd}{\pd N} \int_{-x_{\textrm{B}}}^{x_{\textrm{B}}} \left[ \frac{1}{2} \rho_0 U_0 + P(\rho_0) \right] \, \dd x = -k^2 \delta N \frac{\pd E}{\pd N},
\end{equation}
where
\begin{equation}\label{eqn:Eint}
E = \int_{-x_{\textrm{B}}}^{x_{\textrm{B}}} \left[ \frac{1}{2} \rho_0 U_0 + P(\rho_0) \right] \, \dd x.
\end{equation}
Thus we see that the equilibrium is unstable if $\partial E/\partial N$ is negative.  For the case $\gamma = 2$, we can analytically obtain $E$ and show that $\partial E/\partial N < 0$ by using explicit forms for $\rho_0(x)$ (\Eq{eqn:den}) and $U_0(x)$ (see \myS \ref{sec:quadraticpressure}) and inserting them in \Eq{eqn:Eint}.  In addition, our numerical calculations of $E$ as a function of $N$ for pressures $P = c \rho^{\gamma}$ show that $\partial E/\partial N < 0$ for all $\gamma$ where finite equilibria exist ($\gamma > 1$).  See, for example, Fig.\ \ref{fig:8} for plots of $E(N)$ for $\gamma = 1.1$, 1.4, 1.7, 3, 5, and 7, as well as $P = \rho^2$ for $K=1.25$, 2, and 5.  These plots show that $E(N)$ decreases monotonically with $N$ in all cases.  Thus we conclude that our planar equilibria are always unstable.

We interpret the basic reason for the long-wavelength instability that we have found as being due to the long-range attraction of flock members.  Since, at long wavelength, attraction cannot be balanced by repulsion, long-wavelength density modulations transverse to $x$ tend to grow.  Eventually, as the density enhancements collapse under the attractive force to smaller spatial size, we expect that the repulsion will come into play, and an equilibrium where attraction and repulsion balance in all directions will be established.  Our three-dimensional ellipsoidal equilibrium (\myS \ref{sec:3D}) is of this type and is thus expected not to be susceptible to this instability.  Our planar analysis (Secs.\ \ref{sec:flockcond}, \ref{sec:flockedge}, \ref{sec:quadraticpressure}, and \ref{sec:1D}) has been useful in gaining understanding and in formulating our three-dimensional ellipsoidal equilibria (\myS \ref{sec:3D}).  However, based on the above discussion, we expect only our three-dimensional equilibria (and not our planar equilibria) to be relevant in modeling real situations.

While the discussion of stability in the previous paragraph suggests that the particular mechanism causing long-wavelength instability of the sheet equilibrium may be absent in our elliptical equilibrium, we caution that other, shorter-scale instability mechanisms are possible. For example, numerical results from an agent based model (Ref.\ \cite{Gregoire03}), suggest that interactions tending to equalize the alignment of flock members may be necessary for stability of a finite group. This kind of interaction would not affect our equilibria, which have all flock members aligned, but might conceivably be necessary for stability of our ellipsoidal equilibrium. This issue deserves further study (e.g., one simple way of including alignment interaction in our fluid-like formulation is given in Ref.\ \cite{Mecholsky2010_OA}).  Another issue that we have not addressed and that is open for further study is that of stability in the case of topological interactions (Sec.\ \ref{sec:topology}).

\section{Conclusions}\label{conclusions}
In this paper we have used a dynamic continuum model to investigate animal flocking.  We model the short-range repulsion of flock members by a pressure-like term, $P=c \rho^{\gamma}$.  Adopting the hypothesis that `reasonable equilibria' go to zero flock density at a well-defined flock boundary, we find that such equilibria only occur if $\gamma > 1$, and we investigate the form of these equilibria.  Adopting the further reasonableness hypothesis that, as the spatial size of the flock gets larger and larger, the interior density remains bounded, we find that $\gamma$ must exceed 2.  Furthermore, for $\gamma > 2$ we find that large flocks have an approximately constant interior density.  

We considered planar flocks depending only on one Cartesian coordinate as well as ellipsoidal shaped flocks.  The analysis of planar flocks was analytically convenient because the basic nonlinear equilibrium equation could be integrated by the method of quadrature.  Insight gained from the analysis of planar flocks allowed a good understanding of ellipsoidal flocks, particularly in the case of large flocks with $\gamma > 2$. 

Our ellipsoidal flocks can be thought of as resulting from anisotropic sensing and response of flocking individuals to their neighbors.  In our modeling scheme this resulted in an ellipsoidal shape with principle axes aspect ratios that were independent of flock size, and this feature is in apparent agreement with observation \cite{Ballerini2008201}. 

Recent work \cite{Ballerini2008PNAS} has suggested that an individual in a flock interacts with an approximately fixed number of other flock members.  Thus the geometrical interaction range is larger/smaller when the flock density is lower/higher.  Reference \cite{Ballerini2008PNAS} calls this behavior interaction through `topological distance'.  We have shown that our model and results based on purely-geometrical distance-determined interactions can be simply transformed into a model and corresponding results incorporating topological distance.  Thus our main conclusions are robust in that they apply for geometrical distance and topological distance interaction models.

Finally we investigated the long-wavelength stability of our planar equilibria and found that they were always unstable to perturbations within the plane.  We interpret this instability as being due to the domination of long-range attraction over repulsion at long wavelength.  Thus, of our two types of equilibria solutions (planar and ellipsoidal), we believe that the ellipsoidal type is more relevant for modeling.

This work was supported by ONR grant N00014-07-1-0734.  We dedicate this paper to the memory of our coauthor, Parvez Guzdar, who passed away recently.

%\bibliography{./../../../BIBTEX/flocking}

\end{document}